\documentclass[10pt, conference]{IEEEtran}
\usepackage{algorithmicx}
\usepackage[ruled,vlined,linesnumbered]{algorithm2e}
\usepackage{graphicx}
\usepackage{makecell}
\usepackage{subcaption}
\usepackage{booktabs}
\usepackage{multirow}
\usepackage{hhline}
\usepackage{amsmath,mathtools}
\usepackage{amsfonts,amssymb}
\usepackage{mathrsfs}
\usepackage{gensymb} % for degree
\usepackage{caption}% http://ctan.org/pkg/caption
\usepackage{multirow}
\usepackage{graphicx} 
\usepackage{multirow}
\usepackage{enumitem,color}
\usepackage{algpseudocode}
\usepackage{caption}
\usepackage{titlesec}
\titlespacing*{\section}{1pt}{0.25ex}{0.25ex}
\titlespacing*{\subsection}{1pt}{0.25ex}{0.25ex}
\titlespacing*{\subsubsection}{1pt}{0.25ex}{0.25ex}

\begin{document}
\title{\huge \emph{RadTwin}: Generalizable Wireless Digital Twin for Dynamic Environments \vspace{-0.1in}}

% \title{\huge RadTwin: A Digital Network Twin Framework for Wireless Radiance Field Reconstruction in Dynamic Scenes}

\author{
\IEEEauthorblockN{
Yuru Zhang\IEEEauthorrefmark{1}, 
Ming Zhao\IEEEauthorrefmark{1},
Qiang Liu\IEEEauthorrefmark{1},
Ahmed Alkhateeb\IEEEauthorrefmark{2},
Abhishek K. Agrawal\IEEEauthorrefmark{3},
Qi Qu\IEEEauthorrefmark{3}\\
\IEEEauthorrefmark{1} University of Nebraska-Lincoln,
\IEEEauthorrefmark{2} Arizona State University,
\IEEEauthorrefmark{3} Meta Platforms Inc.,
\\
\IEEEauthorrefmark{1} \{yzhang176, mzhao7, qiang.liu\}@nebraska.edu,
\IEEEauthorrefmark{2} alkhateeb@asu.edu,
\IEEEauthorrefmark{3} \{abhishekag, qqu\}@meta.com
\vspace{-0.2in}
}
% \IEEEauthorblockN{Yuru Zhang, Ming Zhao, Qiang Liu \vspace{-0.16in}}\\
% \IEEEauthorblockA{
% University of Nebraska-Lincoln\\
% yzhang176@huskers.unl.edu}\vspace{-0.4in}
% \and
% \IEEEauthorblockN{Abhishek Agrawal, Qi Qu \vspace{-0.16in}}\\
% \IEEEauthorblockA{
% Meta Platforms Inc.\\
% \{abhishekag, qqu\}@meta.com}\vspace{-0.4in}
}

\maketitle

% \begin{abstract}
% Precisely modeling radio propagation in wireless network has been a significant challenge, especially with the dynamically movement object in the scene. Traditional method, such as ray tracing or neural-based model, often fail to represent the channel in dynamic environments. To solve the problem, we propose RadTwin, a novel framework for wireless channel modeling in dynamic environment.
% \end{abstract}

\begin{abstract}
Precisely modeling radio propagation in dynamic wireless environments is fundamental to the realization of wireless digital twins. 
Traditional ray tracing methods rely on accurate 3D models with detailed environment parameters, while recent neural radiance field approaches learn representations tied to specific static scenes, requiring retraining when environments change.
In this paper, we propose RadTwin, a generalizable wireless digital twin framework that explicitly conditions on scene geometry, enabling adaptation to dynamic environments without retraining. 
RadTwin comprises three key components: 1) a scenario representation network that extracts high-level latent scene features from point clouds, 2) an electromagnetic ray tracing module that computes physics-informed sparse attention masks identifying voxels that physically contribute signals toward each query direction, and 3) a neural propagation decoder that aggregates relevant scene features through masked cross-attention to learn how radio propagation behaves within the given scene geometry.
We evaluate RadTwin on a customized dataset of indoor scenes with varying furniture arrangements. 
Experimental results show that RadTwin achieves 31.6\% higher SSIM (0.846 vs. 0.643) and 91.96\% lower LPIPS (0.023 vs. 0.286) compared to NeRF$^2$.
RadTwin further demonstrates superior cross-scale performance and high generalization and data efficiency, representing a significant advancement toward practical digital network twins for dynamic wireless environments.
\end{abstract}


%==============================================================================
% Introduction Section
%==============================================================================

\section{Introduction}
\label{sec:introduction}

% digital radio twin --> facilitate network optimization, etc.

% example, 

% digital twin attributes, fidelity, synchronicity, scabalbile, see oneTwin paper introduction,

% The rapid evolution of wireless communication systems demands accurate and efficient channel modeling to support network planning, resource allocation, and real-time optimization. Digital network twin (DNT), which creates a virtual replica of the physical wireless environment, has emerged as a promising paradigm for predicting signal propagation characteristics~\cite{wu2021digital}. By maintaining a synchronized digital representation of the radio environment, DNT enables a wide range of applications including coverage optimization, interference management, beam tracking, and predictive handover in next-generation networks.

Digital network twin (DNT) has emerged as a transformative paradigm for next-generation wireless systems~\cite{khan2022digital}, creating virtual replicas of physical radio environments that can be queried for channel prediction and network optimization~\cite{wu2021digital,poorzare2025network}. An effective DNT must satisfy three essential attributes: fidelity, accurately replicating real-world propagation characteristics; synchronicity, tracking environmental changes in a timely manner; and scalability, efficiently adapting to diverse deployment scenarios without prohibitive overhead~\cite{almasan2022network}. These attributes enable a wide range of applications including coverage optimization, beam tracking, interference management, and predictive resource allocation in 5G-and-beyond networks~\cite{tran2025digital,liu2024digital,wang2024digital}. At the core of DNT lies accurate wireless channel modeling, which captures how signals propagate through the physical environment.

\begin{figure}[!t]
	\centering
	\includegraphics[width=3.45in]{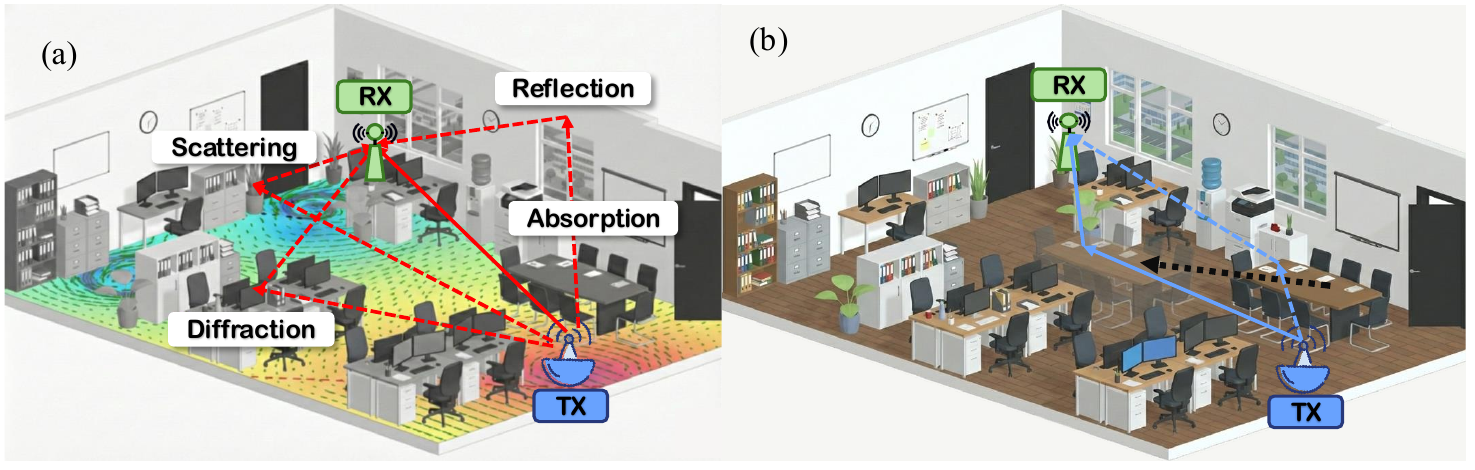}
	\caption{\textbf{Impact of dynamic scene changes on radio propagation.} (a) RF radiance field showing multipath effects (e.g., reflection, scattering, diffraction, and absorption) caused by obstacles. (b) Indoor environment with furniture rearrangement, where the office desk moves along the indicated trajectory. Object movement induces variations in the radiance field.}
	\label{fig:RF_dynamic}
\end{figure}
% traditional ray-tracing based approach, 
% subject to highly accurate 3d modeling with EM property, e.g., material, which is not applicable in frequenctly changed or dynamic scenarios, especially dynamic enviornment
% 

% Traditional wireless channel modeling relies on ray tracing methods to simulate electromagnetic wave propagation by tracing ray paths through the environment and computing interactions with obstacles~\cite{yun2015ray}. While ray tracing provides physically accurate predictions, it is inherently computationally intensive and typically performed offline. More critically, conventional ray tracing struggles to adapt to dynamic changes in the environment, when furniture is rearranged or obstacles are moved, the entire simulation must be re-executed. This limitation prevents ray tracing from providing timely channel state updates essential for real-time wireless applications.

Traditional wireless channel modeling relies on deterministic ray tracing methods that simulate electromagnetic (EM) wave propagation by tracing ray paths and computing interactions with environmental obstacles~\cite{yun2015ray}. As shown in Fig.~\ref{fig:RF_dynamic}(a), Radio Frequency (RF) signals interact with obstacles in the scene through complex physical phenomena including reflection, scattering, diffraction, and absorption, making accurate EM ray tracing computationally prohibitive. 
Moreover, ray tracing requires highly accurate 3D scene models with detailed EM properties (e.g., material permittivity and conductivity)~\cite{an2025radiotwin} for each obstacle. This dependency on precise geometric and material specifications makes ray tracing impractical beyond simulation. As shown in Fig.~\ref{fig:RF_dynamic}(b), when furniture is rearranged, the propagation paths between TX and RX are changed, which causes significant variations in the radiance field. This necessitates 3D model reconstruction and material property re-calibration, which demands substantial manual effort and prevents timely channel state updates essential for real-time wireless applications.

% recent advances in nerual radiance field, (how it works) (what is itsgood side, advantage)
% however, then limitations (e..g, RF-3DGS papers), e.g., synchrinicity to dynamic scenes

% Recent advances in neural radiance fields have inspired a new class of learning-based channel models that encode propagation characteristics into neural network weights~\cite{zhao2023nerf2, lu2024newrf}. These methods achieve impressive accuracy within trained scenes by learning implicit representations of the radio environment. However, they share a fundamental limitation: the scene geometry is encoded implicitly within the network parameters, necessitating complete retraining whenever the environment changes. This per-scene training paradigm is impractical for real-world deployment where indoor environments are inherently dynamic and diverse.

Recent advances in neural radiance fields (NeRF)~\cite{mildenhall2021nerf} have inspired learning-based channel models that encode propagation characteristics into neural network weights~\cite{zhao2023nerf2, lu2024newrf}. These methods represent scenes as continuous volumetric functions, where a neural network learns to predict signal properties at arbitrary spatial locations after training with sparse measurements. NeRF-based approaches achieve impressive prediction accuracy within trained scenes and offer faster inference than ray tracing. However, they share a fundamental limitation: the scene geometry is encoded implicitly within network parameters, necessitating complete model retraining whenever the environment changes. This per-scene training paradigm conflicts with the synchronicity requirement of DNTs, as indoor environments are inherently dynamic with frequent furniture rearrangements and object movements.

In this paper, we propose \emph{RadTwin}, a novel digital radio twin framework designed for wireless channel prediction in dynamic scenes. 
Our key insight is that explicitly conditioning on scene geometry, rather than encoding it implicitly, enables rapid adaptation to environmental changes without retraining. 
We leverage point clouds as the geometric representation, which can be efficiently captured and updated through LiDAR sensors or cameras. 
RadTwin processes geometric input through a scenario representation network that extracts hierarchical voxel features, computes physics-informed sparse attention masks via Line-of-Sight (LOS) visibility, and aggregates relevant features through a Transformer-based decoder to predict spatial spectrum at arbitrary positions. 
This design explicitly separates scene representation from propagation learning, enabling a one-time trained model to generalize across diverse indoor configurations.

% Paragraph 5: Evaluation summary
We evaluate RadTwin on 30 indoor scenes with varying furniture arrangements generated using the Sionna RT simulator. 
RadTwin achieves superior spatial spectrum prediction accuracy on held-out test scenes, with a median SSIM of 0.846 and LPIPS of 0.023, substantially outperforming NeRF$^2$ (SSIM 0.643, LPIPS 0.286) and standard multi-layer perceptron (MLP) baselines. 
The results validate that explicit geometric conditioning enables effective adaptation to scenario changes, representing a significant step toward practical wireless digital twins for dynamic environments. 

%\textcolor{red}{key result}

% Contributions
Overall, we propose RadTwin as a novel generalizable DNT framework for dynamic wireless environments. The main contributions are summarized as follows:

\begin{itemize}
    \item We design a new neural network architecture that explicitly conditions on scene geometry, enabling generalization to dynamic environment configurations without scene-specific retraining. %This addresses the synchronicity challenge faced by existing implicit neural representations.
    
    \item We design a physics-informed sparse attention mechanism, which guides the model to focus on geometrically relevant regions, improving both prediction accuracy and computational efficiency.
    
    \item We evaluate RadTwin with a customized dataset of 30 indoor dynamic scenes, and the results demonstrate superior prediction performance to state-of-the-art solutions.
\end{itemize}

% The remainder of this paper is organized as follows...

%  talk about the problem of wireless channel modelling
% talk about radio radiance field (see NERF2 and RF-3DGS, etc.)

\section{Fundamentals of Radio Radiance Field}
\label{sec:preliminaries}
In this section, we introduce the fundamental concepts of wireless channel modeling and radio radiance fields.

\subsection{Wireless Channel Model}
\label{subsec:channel_model}
A wireless communication system consists of a transmitter (TX) that generates and modulates a signal, which propagates through the wireless channel to a receiver (RX). The transmitted signal can be represented as a complex number $X = A e^{j\psi}$, where $A$ and $\psi$ denote the amplitude and phase, respectively. In the simplest case of a single propagation path, the received signal undergoes amplitude attenuation and phase rotation:
\begin{equation}
    Y = X \cdot \Delta A \, e^{j\Delta \psi} = A \cdot \Delta A \, e^{j(\psi + \Delta \psi)},
    \label{eq:single_path}
\end{equation}
where $\Delta A$ and $\Delta \psi$ denote the amplitude attenuation and phase rotation incurred during propagation.

In real-world environments, EM waves undergo complex interactions with obstacles, including reflection, diffraction, refraction, and scattering. Consequently, multiple signal copies arrive at the RX via different propagation paths. The received signal can be modeled as a superposition of $M$ multipath components:
\begin{equation}
    Y = A \sum_{m=0}^{M-1} \Delta A_m \, e^{j(\psi + \Delta \psi_m)},
    \label{eq:multipath_channel}
\end{equation}
where $\Delta A_m$ and $\Delta \psi_m$ denote the amplitude attenuation and phase shift of the $m$-th path, respectively.

\begin{figure}[t]
  \centering
  \begin{subfigure}[t]{0.135\textwidth}
    \centering
    \includegraphics[width=\linewidth]{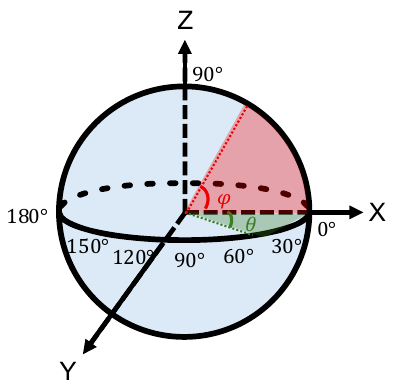}
    \caption{Azimuth \& Elevation}
    \label{fig:ss_a}
  \end{subfigure}
  \hfill
  \begin{subfigure}[t]{0.175\textwidth}
    \centering
    \includegraphics[width=\linewidth]{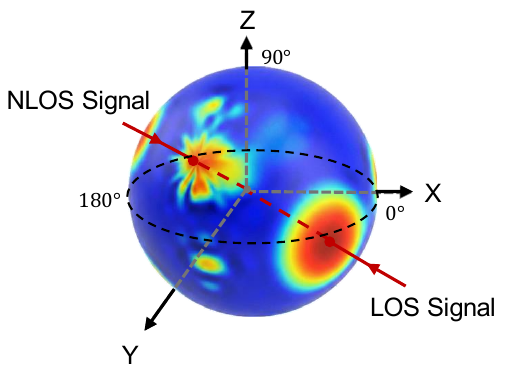}
    \caption{3D Spatial Spectrum}
    \label{fig:ss_b}
  \end{subfigure}
  \hfill
  \begin{subfigure}[t]{0.15\textwidth}
    \centering
    \includegraphics[width=\linewidth]{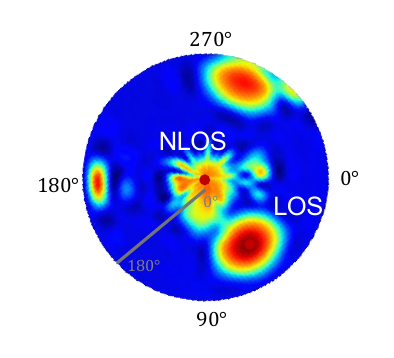}
    \caption{2D Projection}
    \label{fig:ss_c}
  \end{subfigure}

  \caption{\textbf{Illustration of spatial spectrum.}}
  \label{fig:spatial_spectrum}
\end{figure}

When the RX employs a directional antenna, it can selectively receive signals from a specific direction $d=(\theta, \varphi)$, where $\theta \in [0^\circ, 360^\circ)$ denotes the azimuth angle and $\varphi \in [0^\circ, 180^\circ]$ denotes the elevation angle, as shown in Fig.~\ref{fig:spatial_spectrum}(a). We denote the received signal from direction $d$ as $Y(\theta, \varphi)$. By measuring the received power across all angular directions, we obtain the spatial spectrum $\Psi(\theta, \varphi)$, which quantifies the power distribution over the angular domain:
\begin{equation}
    \Psi(\theta, \varphi) = \left| Y(\theta, \varphi) \right|^2.
    \label{eq:spatial_spectrum}
\end{equation}

The spatial spectrum can be viewed as a 2D heatmap showing the power distribution across $N$ angular directions. With one-degree resolution where $\theta \in \{0^\circ, 1^\circ, \ldots, 359^\circ\}$ and $\varphi \in \{0^\circ, 1^\circ, \ldots, 180^\circ\}$, we have $N = 360 \times 181$ directions, and the spatial spectrum forms a matrix:
\begin{equation}
    \Psi = \begin{pmatrix}
        \Psi(0^\circ, 0^\circ) & \Psi(1^\circ, 0^\circ) & \cdots & \Psi(359^\circ, 0^\circ) \\
        \Psi(0^\circ, 1^\circ) & \Psi(1^\circ, 1^\circ) & \cdots & \Psi(359^\circ, 1^\circ) \\
        \vdots & \vdots & \ddots & \vdots \\
        \Psi(0^\circ, 180^\circ) & \Psi(1^\circ, 180^\circ) & \cdots & \Psi(359^\circ, 180^\circ)
    \end{pmatrix}.
    \label{eq:spectrum_matrix}
\end{equation}

This spatial spectrum matrix comprehensively characterizes the multipath propagation environment, revealing dominant propagation paths and their angular distributions. The path loss $L(\theta, \varphi)$ in decibels for each direction can be derived as:
\begin{equation}
    L(\theta, \varphi) = -10\log_{10}\left(\Psi(\theta, \varphi)\right) \; \text{[dB]}.
    \label{eq:path_loss}
\end{equation}
Fig.~\ref{fig:spatial_spectrum}(b) shows the spatial spectrum in 3D, while Fig.~\ref{fig:spatial_spectrum}(c) illustrates the corresponding 2D projection on the X-Y plane.

\begin{figure*}[!t]
	\centering
	\includegraphics[width=6in]{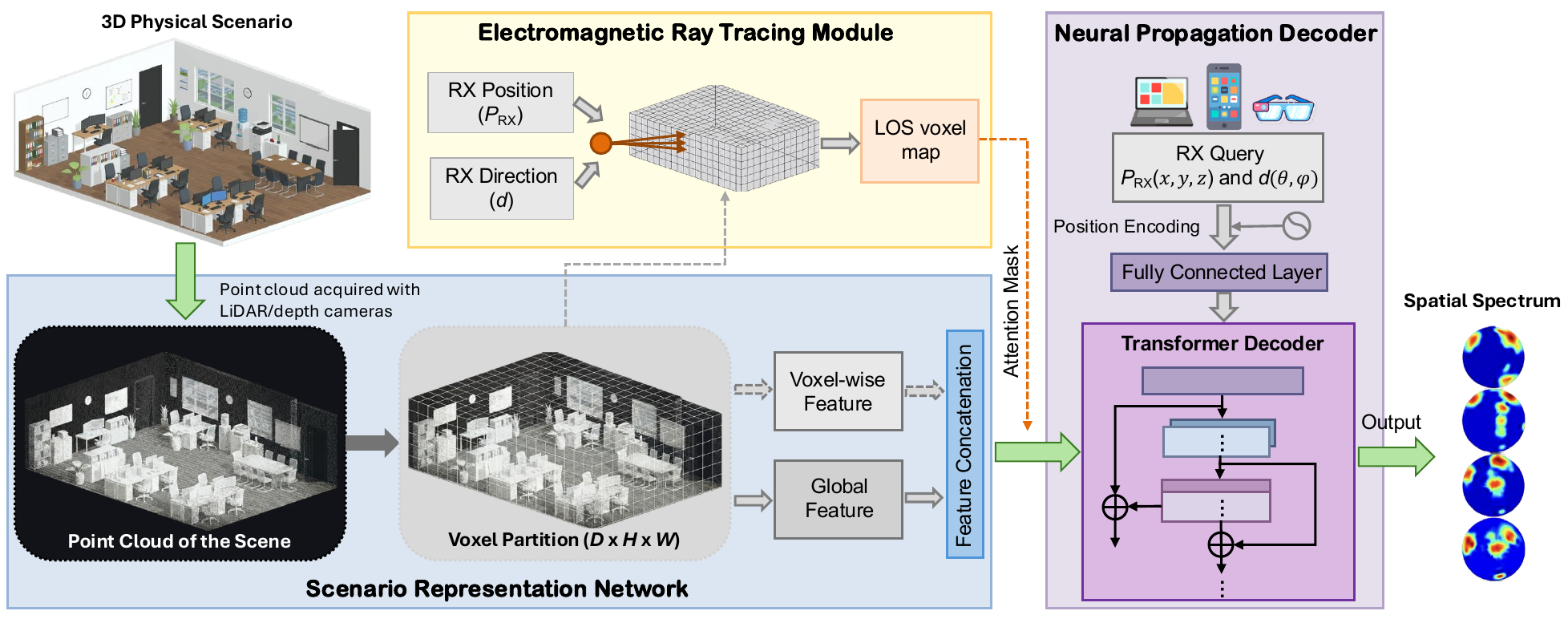}
	\caption{\textbf{An overview of the RadTwin framework.} RadTwin consists of three main components. The point cloud of a 3D scene is first processed by the scenario representation network, which partitions space into a voxel grid and extracts voxel-wise and global features. Given an RX query, the electromagnetic ray tracing module computes LOS voxel maps to generate sparse attention masks. Finally, the neural propagation decoder aggregates relevant voxel features through a Transformer decoder with masked cross-attention to synthesize the spatial spectrum.} 
	\label{fig:framework}
\end{figure*}

\subsection{Radio Radiance Field}
\label{subsec:radio_radiance_field}
In this subsection, we introduce the concept of radio radiance field (RRF), which provides the theoretical foundation for existing neural channel modeling approaches. 

The RRF provides a continuous representation of EM wave propagation in a given environment. Analogous to the optical radiance field in computer vision, which maps a 3D position and viewing direction to color and density, the RRF characterizes the wireless channel as a function of TX and RX configurations. Unlike the EM field where almost all spatial points possess a well-defined field vector, the RRF describes radiance only at object surfaces where EM waves undergo interactions such as reflection, diffraction, and scattering.

A practical RRF representation requires modeling both scene geometry and the radiance function. The geometry is characterized by a density field $\alpha(P_x)$, where $P_x \in \mathbb{R}^3$ denotes a continuous 3D position. This density takes high values for solid objects, low values for translucent materials, and zero for free space. The radiance function $c(P_x, d)$ describes the signal emitted from position $P_x$ toward direction $d = (\theta, \varphi)$. For a given TX at position $P_{\text{TX}} \in \mathbb{R}^3$, the complete RRF can be expressed as:
\begin{equation}
    \mathcal{R}(P_{\text{TX}}) = \{c(P_x, d), \alpha(P_x)\}.
    \label{eq:rrf_representation}
\end{equation}
% This formulation enables querying the spatial spectrum $\Psi$ at any arbitrary RX position $P_{\text{RX}}$ within the scene.

NeRF$^2$~\cite{zhao2023nerf2} pioneered the application of neural radiance fields to wireless channel modeling by parameterizing the RRF using MLPs. It treats each voxel at position $P_x$ as a virtual retransmitter that combines and retransmits signals received from all possible paths. The neural network learns to predict the radiance field $f$ as:
\begin{equation}
    f_\Theta: (P_{\text{TX}}, P_x, d) \Rightarrow \left(\delta(P_x), \mathcal{S}(P_x, d)\right),
    \label{eq:nerf2}
\end{equation}
where $\delta(P_x)$ represents the attenuation coefficient determined by material properties, and $\mathcal{S}(P_x, d)$ is the directional signal retransmitted toward direction $d$. The final received signal is obtained by integrating contributions from all voxels along each ray using volume rendering. While this approach demonstrates the feasibility of neural representations for wireless channels, it requires per-scene training as the scene geometry is implicitly encoded within the network parameters $\Theta$.

\section{RadTwin Overview}
\label{sec:overview}
In this section, we introduce the RadTwin, a novel wireless digital twin framework designed for generalizable radio propagation modeling in dynamic environments. 
Unlike existing approaches that require per-scene training, RadTwin learns scene-agnostic propagation patterns from explicit geometric representations, enabling generalization to dynamic scenarios.

% then talk about the System Overview
\textbf{Problem.}
We consider a wireless communication scenario where a TX is deployed at a fixed position and an RX is located at position $P_{\text{RX}} \in \mathbb{R}^3$ within the environment. For each RX position, we query the received signal along a specific direction $d = (\theta, \varphi)$, where $\theta$ and $\varphi$ denote the azimuth and elevation angles, respectively. Due to multipath propagation, the signal received from a given direction is the superposition of multiple paths that undergo reflection, diffraction, and scattering from surrounding obstacles. The environment geometry, denoted as $E$, encompasses all objects such as walls, floors, and furniture, whose spatial configuration significantly influences the radio propagation characteristics.

Given the environment geometry $E$ and RX query $(P_{\text{RX}}, d)$, RadTwin predicts the channel metric $S$ (e.g., RSRP and RSSI) along all propagation paths arriving from direction $d$. The learned model generalizes across different environment configurations without scene-specific retraining. This is formulated as $f_\Theta: (E, P_{\text{RX}}, d) \Rightarrow S$, where $\Theta$ represents the learnable parameters of the network. 
Overall, the model is trained to minimize the mean squared error between predicted and ground-truth channel metrics using the following loss function:
\begin{equation}
    \mathcal{L} = \frac{1}{|\mathcal{D}|} \sum_{i=1}^{|\mathcal{D}|} \left(\hat{S}_i - S_i\right)^2,
    \label{eq:loss}
\end{equation}
where $\mathcal{D} = \{(E_i, P_{\text{RX},i}, d_i, S_i)\}_{i=1}^{|\mathcal{D}|}$ denotes the training dataset comprising samples from multiple scenes, and $\hat{S}_i = f_\Theta(E_i, P_{\text{RX},i}, d_i)$ is the predicted channel metric.

\textbf{Overview.}
As shown in Fig.~\ref{fig:framework}, given the 3D environment geometry and an RX query specifying position and direction, RadTwin predicts the corresponding signal. 
The framework is built upon a key insight: wireless signal propagation is fundamentally governed by the geometric structure of the environment, particularly the spatial arrangement of obstacles along the propagation path. 
Hence, we architect RadTwin with three key components:

\begin{itemize}[leftmargin=*]
    \item \textbf{Scenario Representation Network}: This module transforms raw environment geometry into a structured voxel-based representation. The network extracts both local geometric features capturing fine-grained obstacle properties and global contextual features encoding the overall spatial layout through 3D convolutions.

    \item \textbf{Electromagnetic Ray Tracing Module}: This module identifies geometrically relevant regions for each query direction based on physical propagation principles. We compute the intersections between the query rays and the voxel grid using ray-box intersection tests, producing sparse LOS voxel indices that encode propagation constraints.
    
    \item \textbf{Neural Propagation Decoder}: This module synthesizes signal predictions by aggregating information from relevant voxels. A Transformer decoder takes the RX query and attends to voxel features through cross-attention, constrained by the LOS masks. This physics-informed attention mechanism guides the model toward physically meaningful feature aggregation.
\end{itemize}

The RadTwin framework introduces a novel approach to neural radio propagation modeling that achieves generalization across diverse environment configurations. The key innovation of RadTwin lies in its explicit geometric conditioning. Unlike implicit neural radiance fields that encode scene geometry within network weights, RadTwin treats the environment as an explicit input. This design yields two benefits: generalization across scenes, where a one-time trained model applies to dynamic configurations without retraining; and interpretability, where LOS-based attention reveals which spatial regions influence each prediction. Together, these components form an end-to-end differentiable pipeline that learns generalizable propagation patterns for dynamic environments.

% The three components work in concert: the scenario representation network extracts hierarchical voxel features combining local and global information; the electromagnetic ray tracing module computes LOS masks encoding physical propagation constraints; and the neural propagation decoder aggregates relevant features through masked cross-attention.

\section{RadTwin Architecture}
\label{sec:architecture}
In this section, we present the detailed architecture of RadTwin's three core components.

\subsection{Scenario Representation Network}
\label{subsec:scenario_rep}
The scenario representation network provides a compact and learnable encoding of environment geometry, serving as the foundation for generalizing across diverse scene configurations. The choice of 3D geometry representation is critical for neural network-based radio propagation modeling. Mesh representations, composed of vertices, edges, and faces, are widely used in computational science and engineering. However, they are costly to acquire, requiring dedicated 3D modeling or reconstruction pipelines, and their use in deep learning is limited due to non-differentiable triangle face indices. Although differentiable mesh processing~\cite{li2018differentiable} and mesh-based generalizable channel modeling~\cite{jiang2025learnable} have been explored, these approaches remain computationally expensive and difficult to maintain in dynamic environments.

In contrast, we adopt point clouds as our scene representation due to their low acquisition cost, wide availability, and scalability. Point clouds have been widely adopted in neural network-based research, with applications in 3D surface reconstruction~\cite{park2019deepsdf, choe2021deep}, geometry denoising~\cite{roveri2018pointpronets, rakotosaona2020pointcleannet, luo2021score}, and shape completion~\cite{wen2021pmp, xiang2021snowflakenet}. Exploiting the differentiability of point clouds, we employ the VoxelNet~\cite{zhou2018voxelnet} architecture for geometry encoding, which was originally proposed for LiDAR-based 3D object detection and demonstrates effectiveness in learning discriminative features from sparse and irregular point distributions.

\begin{figure*}[!t]
	\centering
	\includegraphics[width=5in]{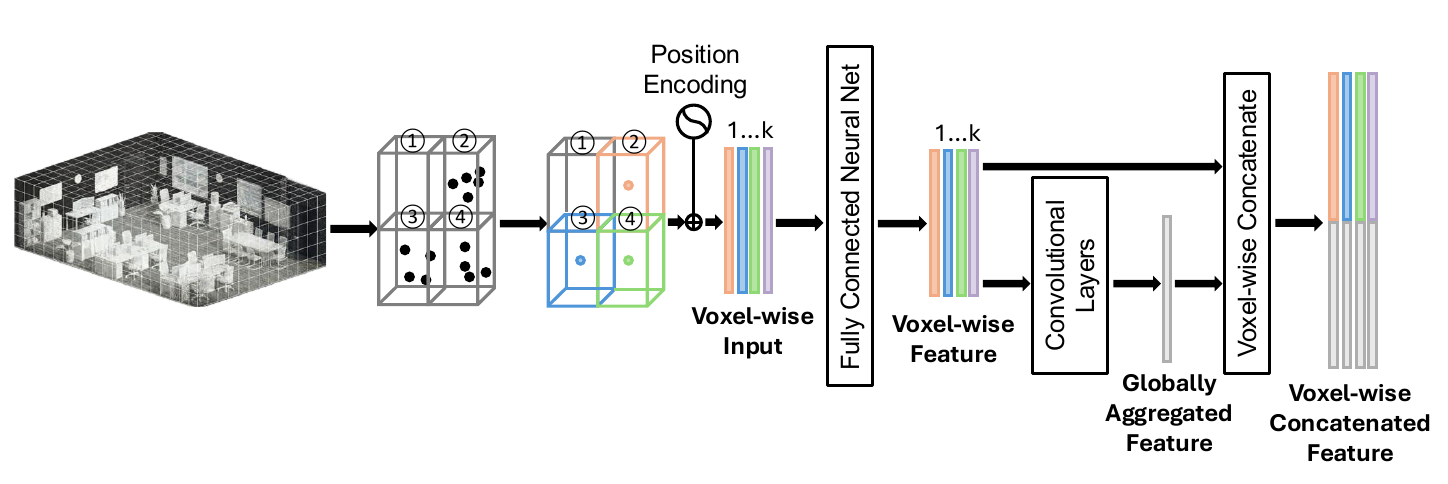}
	\caption{\textbf{Architecture of Scenario Representation Network.} The input point cloud is first partitioned into a voxel grid, where each non-empty voxel is represented by its center coordinates. After positional encoding, a fully connected network extracts voxel-wise local features. These features are then processed by 3D convolutional layers to obtain a globally aggregated feature, which is concatenated with local features to form the final voxel-wise representation.}
	\label{fig:SRN-frame}
\end{figure*}

%The voxelization provides a fixed-dimensional encoding regardless of input complexity and naturally aligns with spatial reasoning for propagation modeling.

The network architecture is shown in Fig.~\ref{fig:SRN-frame}. Given a point cloud encompassing a 3D space with dimensions $D \times H \times W$ along the Z, Y, X axes, we partition the space into voxels of size $v_D \times v_H \times v_W$. The resulting voxel grid has dimensions $D' \times H' \times W'$, where $D' = D/v_D$, $H' = H/v_H$, and $W' = W/v_W$. Points are grouped according to the voxel in which they reside. Due to the sparse distribution of scene geometry, many voxels remain empty. We retain only non-empty voxels containing at least $T$ points to filter out noise and outliers. Each valid voxel is represented by its center coordinates $C_k = (c_x, c_y, c_z) \in \mathbb{R}^3$, yielding $K$ occupied voxels $\{C_k\}_{k=1}^K$.

The voxel centers are transformed through positional encoding before feature extraction. In radio propagation, received signal strength exhibits rapid spatial variations due to multipath interference and small-scale fading, where phase differences of merely half a wavelength can cause significant power fluctuations. Standard neural networks with smooth activation functions struggle to capture such high-frequency variations. Positional encoding addresses this limitation by mapping coordinates into a higher-dimensional space using sinusoidal functions at multiple frequencies:
\begin{equation}
    \footnotesize
    \gamma(P) = \left[\sin(2^0\pi P), \cos(2^0\pi P), \ldots, \sin(2^{L-1}\pi P), \cos(2^{L-1}\pi P)\right],
    \label{eq:pos_enc}
\end{equation}
where $P$ denotes the input coordinate and $L$ controls the number of frequency bands. This encoding enables the network to learn fine-grained spatial dependencies. The encoded voxel centers $\gamma(C_k)$ are processed by a fully connected network to extract local features $f_k \in \mathbb{R}^{d_l}$ for each voxel.

To capture global scene context, we place local features at their corresponding grid positions to construct a dense feature tensor and apply 3D convolutional layers that aggregate information across the entire spatial extent. These convolutional layers progressively expand the receptive field, incorporating contextual information from neighboring voxels. The resulting global feature $\tilde{f} \in \mathbb{R}^{d_g}$ encodes the overall scene structure. Finally, local and global features are concatenated to form the voxel-wise output representation:
\begin{equation}
    f_k^{\text{out}} = [f_k; \tilde{f}] \in \mathbb{R}^{d_l + d_g}, \quad k = 1, \ldots, K.
    \label{eq:voxel_feature}
\end{equation}

By processing only non-empty voxels and representing features as sparse tensors, the network achieves computational efficiency while preserving geometric structure essential for radio propagation prediction.

\subsection{Electromagnetic Ray Tracing}
\label{subsec:ray_tracing}
\begin{figure}[!t]
	\centering
	\includegraphics[width=3.3in]{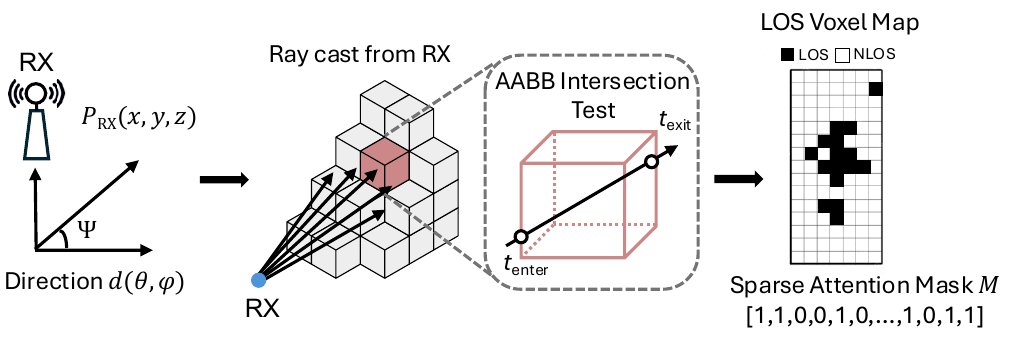}
	\caption{\textbf{Architecture of Electromagnetic Ray Tracing Module.} Given an RX position $P_{\text{RX}}$ and query direction $d$, rays are cast across a discretized spherical grid. For each ray, the AABB intersection test identifies the nearest voxel. The LOS voxels within the angular window centered at $d$ are aggregated to form sparse attention masks.}
	\label{fig:ERT-frame}
\end{figure}

The electromagnetic ray tracing module establishes the physical relationship between RX queries and scene geometry by computing LOS visibility, as shown in Fig.~\ref{fig:ERT-frame}. In radio propagation, signals received from a particular direction predominantly originate from surfaces visible along that direction, and these surfaces represent the final interaction points of potentially complex multipath trajectories. Encoding this physical prior into the neural network architecture improves both prediction accuracy and model interpretability.

For each receiver position $P_{\text{RX}}$, we precompute a LOS voxel map that identifies which voxel is directly visible from each reception direction. The spherical domain is discretized into a grid of $N_\theta \times N_\varphi$ directions. For each direction $d_{ij} = (\theta_i, \varphi_j)$ where $i \in [1, N_\theta]$ and $j \in [1, N_\varphi]$, we convert the spherical coordinates to a unit direction vector:
\begin{equation}
    \vec{d}_{ij} = (\sin\theta_i \cos\varphi_j, \sin\theta_i \sin\varphi_j, \cos\theta_i).
\end{equation}
We then cast a ray from $P_{\text{RX}}$ along $\vec{d}_{ij}$ and determine its intersection with scene voxels using the axis-aligned bounding box (AABB) slab method~\cite{williams2005efficient}. For a voxel with bounding box $[b^{\min}, b^{\max}]$, we compute the ray entry and exit distances along each axis $\alpha \in \{x, y, z\}$:
\begin{equation}
    t_1^\alpha = (b_\alpha^{\min} - p_\alpha) / d_\alpha, \quad t_2^\alpha = (b_\alpha^{\max} - p_\alpha) / d_\alpha,
    \label{eq:ray_slab}
\end{equation}
where $p_\alpha$ and $d_\alpha$ denote the RX position and ray direction components along axis $\alpha$. The overall entry and exit distances are:
\begin{equation}
    t_{\text{enter}} = \max_\alpha \min(t_1^\alpha, t_2^\alpha), \quad t_{\text{exit}} = \min_\alpha \max(t_1^\alpha, t_2^\alpha).
\end{equation}
A valid intersection occurs when $t_{\text{enter}} < t_{\text{exit}}$ and $t_{\text{exit}} > 0$. Among all valid intersections, we record the voxel with the smallest $t_{\text{enter}}$ as the LOS voxel for direction $(\theta_i, \varphi_j)$.

Since the RX query corresponds to a finite angular region rather than an infinitesimal direction, we aggregate LOS voxels within an angular window centered at the query direction $(\theta, \varphi)$:
\begin{equation}
    \mathcal{V}_{\text{LOS}}(\theta, \varphi) = \bigcup_{\substack{\theta_i \in [\theta - \Delta\theta, \theta + \Delta\theta] \\ \varphi_j \in [\varphi - \Delta\varphi, \varphi + \Delta\varphi]}} \mathcal{V}_{\text{hit}}(\theta_i, \varphi_j),
    \label{eq:los_aggregation}
\end{equation}
where $\mathcal{V}_{\text{hit}}(\theta_i, \varphi_j)$ denotes the LOS voxel for direction $(\theta_i, \varphi_j)$, and $\Delta\theta$, $\Delta\varphi$ define the angular window size. This aggregation captures the fact that received signals integrate contributions from a range of angles.

The LOS voxel sets are encoded as sparse attention masks for the neural decoder. For each query, we construct a binary mask $M \in \{0, 1\}^K$ over all $K$ scene voxels:
\begin{equation}
    M_k = 
    \begin{cases}
        0, & \text{if voxel } k \in \mathcal{V}_{\text{LOS}}(\theta, \varphi), \\
        1, & \text{otherwise}.
    \end{cases}
    \label{eq:mask}
\end{equation}
To bound computational cost, we retain at most $N_{\max}$ LOS voxels per query. This sparse masking mechanism serves two purposes: (1) it enforces a physical constraint ensuring predictions are influenced only by geometrically relevant voxels; and (2) it reduces attention complexity from $\mathcal{O}(K)$ to $\mathcal{O}(N_{\max})$, enabling efficient processing of large scenes.

\subsection{Neural Propagation Decoder}
\label{subsec:decoder}
The neural propagation decoder aggregates voxel features under the guidance of LOS masks to predict the received channel metric. The RX query encodes both the spatial position $P_{\text{RX}} = (x, y, z)$ and the reception direction $\vec{d} = (\sin\theta\cos\varphi, \sin\theta\sin\varphi, \cos\theta)$, represented as a Cartesian unit vector. Both components are independently transformed using positional encoding $\gamma(\cdot)$ defined in Eq.~\ref{eq:pos_enc}, concatenated, and linearly projected to form the query embedding:
\begin{equation}
    q_{\text{RX}} = F_{\text{FC}}\left(\left[\gamma(P_{\text{RX}}); \gamma(\vec{d})\right]\right) \in \mathbb{R}^{d_e}.
    \label{eq:rx_query}
\end{equation}
The decoder employs a Transformer architecture where $q_{\text{RX}}$ serves as the query and voxel features $\{f_k^{\text{out}}\}_{k=1}^K$ serve as keys and values. Masked cross-attention is computed as:
\begin{equation}
    \text{Attention}(Q, K, V, M') = \text{softmax}\left(\frac{QK^\top}{\sqrt{d_k}} + M'\right)V,
    \label{eq:attention}
\end{equation}
where $Q = W_Q q_{\text{RX}}$, $K = W_K F$, $V = W_V F$ with $F = [f_1^{\text{out}}, \ldots, f_K^{\text{out}}]^\top \in \mathbb{R}^{K \times (d_l+d_g)}$, $d_k$ is the key dimension, and $M'$ is derived from the binary mask $M$ (Eq.~\ref{eq:mask}) by setting masked positions to $-\infty$, so that non-LOS voxels receive zero attention weight after softmax. The Transformer decoder consists of $N_{\text{layer}}$ layers with multi-head cross-attention, feed-forward sub-layers, residual connections, and layer normalization:
\begin{equation}
    h = F_{\text{TransformerDecoder}}\left(q_{\text{RX}}, \{f_k^{\text{out}}\}_{k=1}^K, M'\right)  \in \mathbb{R}^{d_e}.
\end{equation}
The output $h$ is transformed through a fully connected layer with ReLU activation and clamped to ensure physically plausible predictions:
\begin{equation}
    \hat{S} = \min\left(\text{ReLU}\left(F_{\text{FC}}(h)\right), S_{\max}\right),
    \label{eq:output}
\end{equation}
where $\hat{S}$ denotes the predicted channel metric from direction $d$, and $S_{\max}$ is the upper bound corresponding to the transmitted power.

\section{RadTwin Implementation}
\label{sec:implementation}
In this section, we describe the implementation of RadTwin, including dataset collection and model configuration.
\subsection{Dataset Collection}
\label{subsec:dataset}
\begin{figure}[!t]
	\centering
	\includegraphics[width=3.45in]{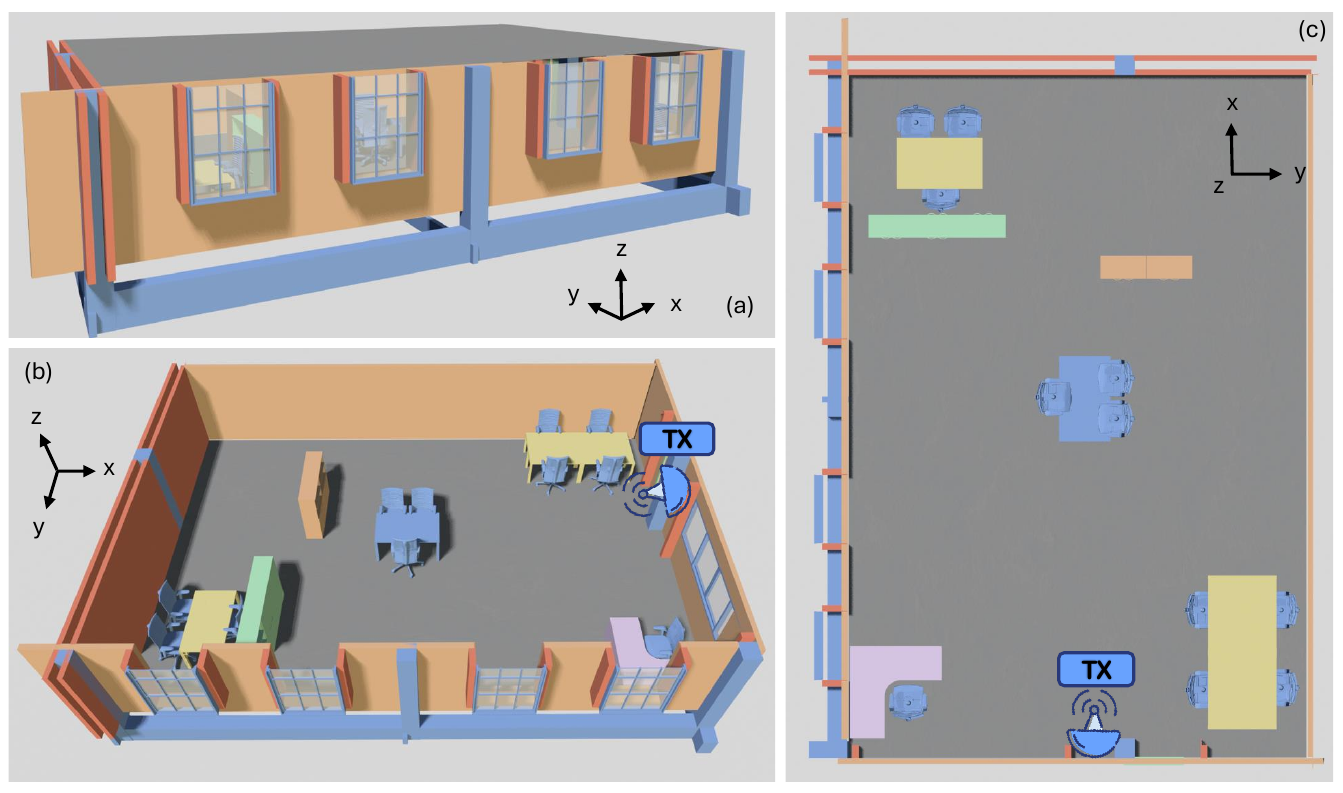}
	\caption{\textbf{Indoor office scene for dataset generation (small size).} (a) Exterior view. (b) Interior 3D view with ceiling removed. (c) Top view.}
	\label{fig:scenario}
\end{figure}

We construct a dataset using Blender 3.0 to evaluate the generalization of RadTwin. The dataset spans three scene sizes: small ($6 \times 4 \times 2.5$\,m$^3$), medium ($12 \times 10 \times 2.5$\,m$^3$), and large ($30 \times 18 \times 2.5$\,m$^3$). For each size, we generate 30 indoor office scenes with identical room dimensions but varying furniture arrangements, where tables, chairs, shelves, and other objects are positioned differently across scenes to simulate real-world environments where object layouts change over time. As shown in Fig.~\ref{fig:scenario}, we illustrate a small scene as an example. For each scene, we sample points uniformly on surfaces of objects (e.g., walls, floors, ceilings, and furniture) to form the point cloud, with the number of points scaling with scene size (e.g., 25,000 for small scenes).

The wireless channel data is generated using Sionna v0.12.0, an open-source ray tracing library. The TX is equipped with an omnidirectional antenna at a fixed position across all scenes, as indicated in Fig.~\ref{fig:scenario}(b). For each scene, we randomly sample 1,000 RX positions within the room volume and compute path loss for all reception directions in $10^\circ$ increments, covering $\theta \in [0^\circ, 350^\circ]$ and $\varphi \in [0^\circ, 180^\circ]$, yielding $36 \times 19 = 684$ directions per RX. The simulation operates at 3.5\,GHz with reflection, refraction, and diffraction enabled.

\begin{figure}[!t] % cannot have space 
\captionsetup{justification=centering}
  \begin{minipage}[t]{0.24\textwidth}
    \centering
    \includegraphics[width=1.7in]{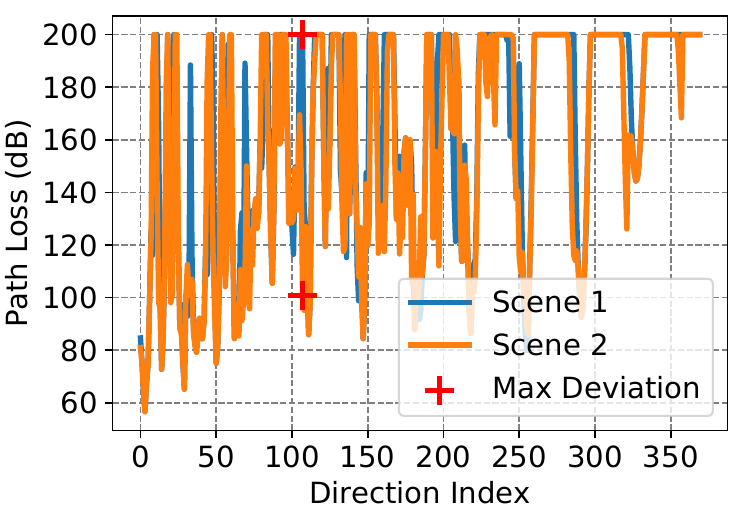}
    \caption{Path loss variation across directions for two small scenes.}
    \label{fig:dataset_direction}
  \end{minipage}
  \begin{minipage}[t]{0.24\textwidth}
    \centering
    \includegraphics[width=1.7in]{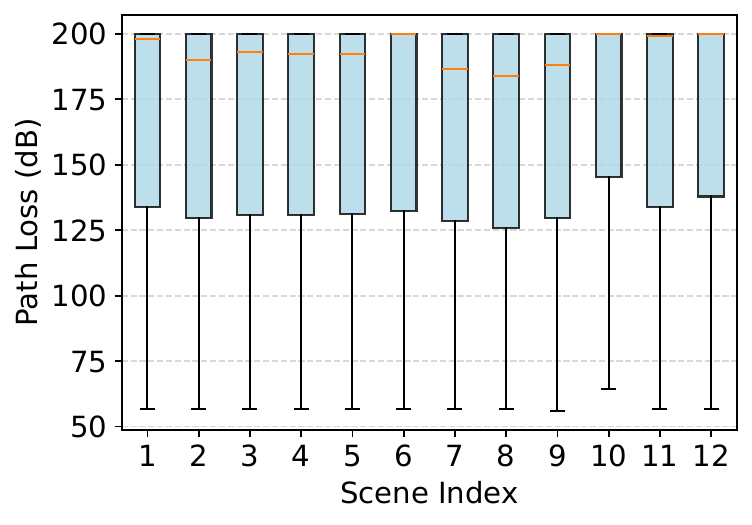}
    \caption{Path loss distribution across 12 small scenes at a fixed RX.}
    \label{fig:dataset_boxplot}
  \end{minipage}
\end{figure}

To quantify the impact of scene variations on radio propagation, we analyze path loss distributions across different configurations using the small scenes. Fig.~\ref{fig:dataset_direction} compares directional path loss at a fixed RX position between two scenes differing only by the displacement of a single bookshelf by 3.5\,m, shown in orange in Fig.~\ref{fig:scenario}. Despite this minor geometric change, certain directions exhibit path loss deviations exceeding 99\,dB, demonstrating the extreme sensitivity of radio propagation to object placement. This is because even small geometric modifications can block or create new propagation paths, fundamentally altering the multipath structure. Fig.~\ref{fig:dataset_boxplot} shows path loss distributions across 12 scenes with varying layouts, where the median and interquartile ranges differ significantly, confirming that layout changes substantially affect channel characteristics.

For each scene size, the dataset is split at the scene level: 24 scenes (80\%) for training and 6 scenes (20\%) for testing. This ensures that test scenes contain furniture configurations not seen during training, providing rigorous evaluation of generalization to dynamic environments.

\subsection{Model Configuration}
\label{subsec:config}
The voxel grid partitions the 3D scene into discrete volumetric cells with a constant voxel size of $0.5 \times 0.5 \times 0.5$\,m$^3$, resulting in grid dimensions that scale with scene size (e.g., $12 \times 8 \times 5$ for small scenes). This resolution balances spatial granularity with computational efficiency. The maximum number of LOS voxels per query is set to $N_{\max} = 16$ to bound memory consumption while preserving the most relevant geometric information.

The RadTwin model is implemented in PyTorch. The scenario representation network extracts a 32-dimensional local feature for each voxel through a fully connected layer, and a 16-dimensional global feature through 3D convolutional layers with channel dimensions progressively increasing from 64 to 256 to 768. Local and global features are concatenated to form the 48-dimensional voxel representation. The neural propagation decoder consists of 3 Transformer decoder layers with single-head attention, a hidden dimension of 128, and a dropout rate of 0.1.

We train the model on an NVIDIA RTX PRO 6000 GPU with 96GB memory using a batch size of 4,096 samples for 30 epochs. We employ the Adam optimizer with lr=0.001 and a step scheduler that reduces the learning rate by a factor of 0.8 every 3 epochs. To ensure memory efficiency, we adopt a scene-aware batch sampler that groups samples from the same scene, allowing voxel features to be computed once and shared across all samples in the batch.

% \textbf{Neural Agent.}
% We implement the neural agent with a BNN model with 4-layer fully connected architecture (i.e., 128x256x256x128), in PyTorch.
% We use the \textit{Tanh} activation function in the BNN \cite{Goodfellow-et-al-2016}. 
% We utilize the \textit{Adadelta} optimizer with the initial learning rate of 0.001, where the learning rate is decayed by using the \textit{StepLR} scheduler with gamma 0.95. 
% % Additionally, the input dimension of the BNN model is 7 (states) and 20 (distribution of network performance), respectively.

\begin{figure*}[!t]
	\centering
	\includegraphics[width=5.5in]{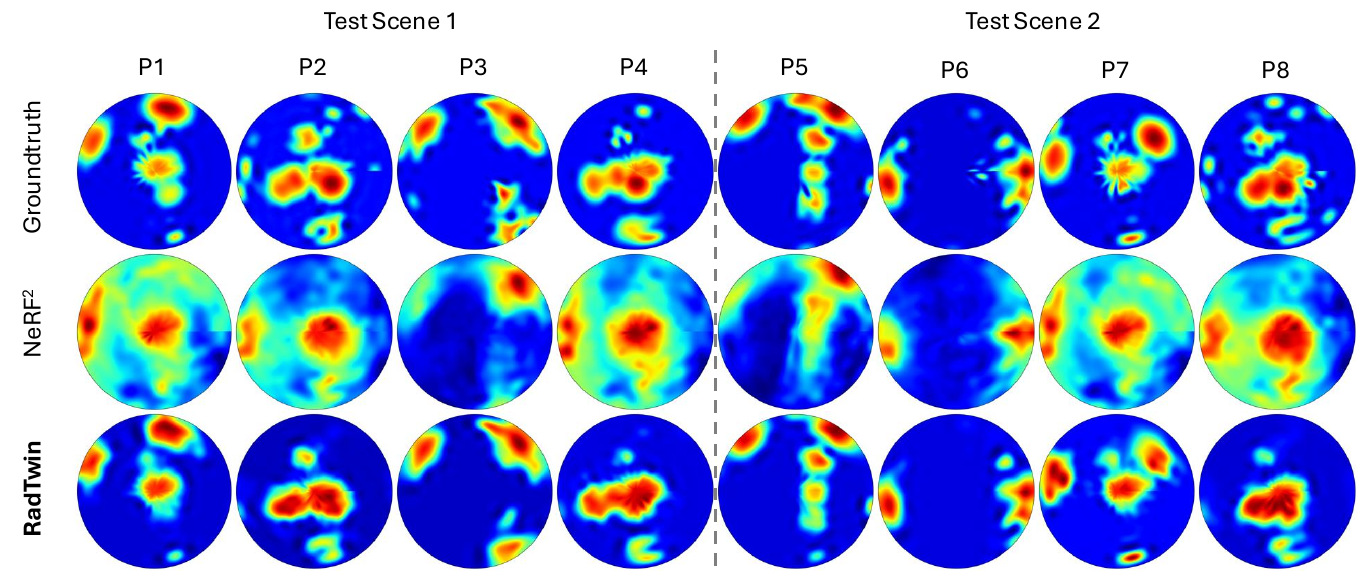}
	\caption{\textbf{Synthesis of spatial spectrums.} Results at eight RX positions across two test scenes. RadTwin accurately captures both dominant propagation paths and fine-grained multipath structures, while NeRF$^2$ exhibits noticeable discrepancies in detailed patterns.}
	\label{fig:spatial_spectrum_new}
\end{figure*}

\section{Performance Evaluation}
\label{sec:evaluation}
In this section, we evaluate RadTwin with state-of-the-art methods on our dataset of dynamic indoor environments. 
% We compare RadTwin against baseline methods and conduct comprehensive experiments to assess its performance under various conditions.

\subsection{Comparison Methods}
\label{subsec:baselines}
We compare RadTwin with the following baseline methods:
\begin{itemize}
    \item \textbf{Ground Truth}: The ground truth spatial spectrum is obtained from Sionna RT simulations. The visualization is generated by converting signal measurements from matrix to polar coordinates. % Ideally, the spectrum should peak in the LOS direction. However, due to multipath propagation in complex environments, multiple peaks may appear corresponding to reflected or diffracted paths.
    
    \item \textbf{NeRF$^2$}~\cite{zhao2023nerf2}: NeRF$^2$ is a neural radiance field-based method for spatial spectrum synthesis. Similar to our work, NeRF$^2$ can synthesize spatial spectra at arbitrary positions after training on a given scene. To match our experimental setup with a fixed transmitter, we modify its implementation to predict spatial spectra at arbitrary RX positions.
    
    \item \textbf{MLP}: A baseline multi-layer perceptron that directly maps RX position and direction to path loss without explicit geometric reasoning. The MLP consists of four fully connected layers with ReLU activations.
\end{itemize}
% Since NeRF$^2$ and MLP implicitly encode scene geometry within their network parameters, they require per-scene training and cannot generalize to dynamic environments. In our cross-scene evaluation, we train separate NeRF$^2$ and MLP models on individual training scenes and evaluate them on test scenes to demonstrate their lack of generalization capability. In contrast, RadTwin is trained on multiple scenes jointly and evaluated on test scenes which is unseen by model.

\subsection{Spatial Spectrum Synthesis}
\label{subsec:spectrum_synthesis}
Fig.~\ref{fig:spatial_spectrum_new} shows a comparison of spatial spectrum synthesis results at eight RX positions across two test scenes. Visually, RadTwin produces spatial spectra that closely match the ground truth, accurately capturing both dominant propagation paths and fine-grained multipath structures. In contrast, NeRF$^2$ captures the general pattern but exhibits noticeable discrepancies in detailed structure, particularly in regions with complex multipath interference.

To quantitatively evaluate synthesis quality, we employ the Structural Similarity Index (SSIM) and Learned Perceptual Image Patch Similarity (LPIPS). SSIM measures structural similarity between two images, with higher values indicating greater similarity. LPIPS evaluates perceptual similarity using deep features, where lower values indicate better quality. We synthesize spatial spectra at 1,000 RX positions across test scenes and compute both metrics for each method.

\begin{figure}[!t] % cannot have space 
\captionsetup{justification=centering}
  \begin{minipage}[t]{0.24\textwidth}
    \centering
    \includegraphics[width=1.7in]{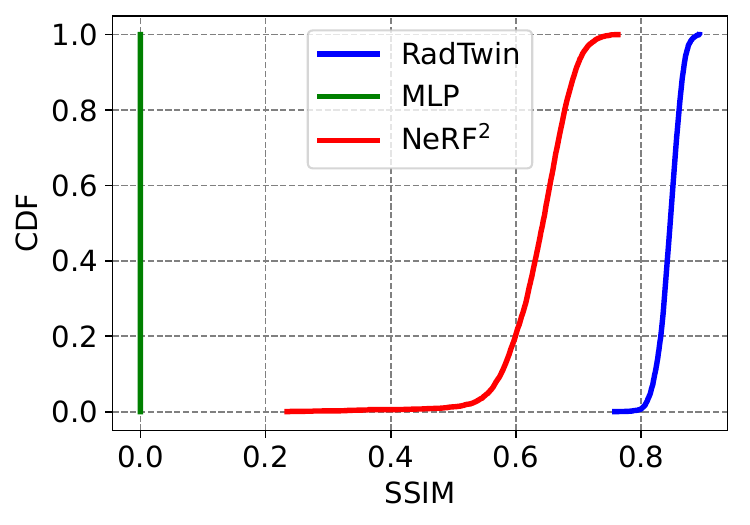}
    \caption{CDF of SSIM values for spatial spectrum synthesis.}
    \label{fig:ssim_cdf}
  \end{minipage}
  \begin{minipage}[t]{0.24\textwidth}
    \centering
    \includegraphics[width=1.7in]{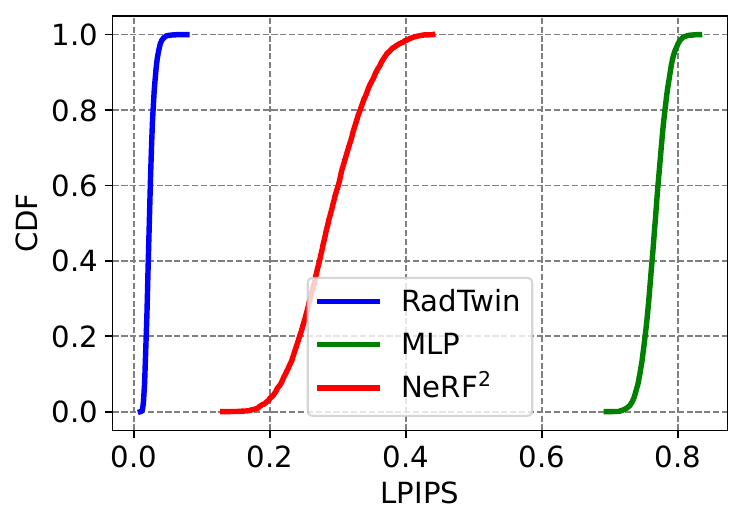}
    \caption{CDF of LPIPS values for spatial spectrum synthesis.}
    \label{fig:lpips_cdf}
  \end{minipage}
\end{figure}

Fig.~\ref{fig:ssim_cdf} shows the cumulative distribution function (CDF) of SSIM values. RadTwin achieves a median SSIM of 0.846 and a 90th percentile of 0.867, significantly outperforming NeRF$^2$ (median 0.643, 90th percentile 0.694). MLP fails entirely with a median SSIM near zero, indicating its inability to synthesize meaningful spatial spectra. Fig.~\ref{fig:lpips_cdf} shows the CDF of LPIPS values, where RadTwin achieves a median of 0.023 and 90th percentile of 0.032, substantially better than NeRF$^2$ (median 0.286, 90th percentile 0.356). MLP exhibits the worst perceptual quality with a median LPIPS of 0.767.

The performance gap can be attributed to fundamental differences in scene representation. MLP lacks explicit geometric reasoning and relies solely on coordinate-based mapping, making it unable to generalize to different scene configurations. NeRF$^2$ incorporates ray-based volume rendering but encodes scene geometry implicitly within network weights, limiting its adaptability to new environments. In contrast, RadTwin explicitly conditions on scene geometry through point cloud input, enabling effective generalization to different furniture arrangements. Furthermore, the physics-informed sparse attention mechanism guides the model to focus on geometrically relevant voxels, improving both accuracy and interpretability.

Table~\ref{tab:time_comparison} compares the computational time of all methods. RadTwin requires 88.8 minutes to train a single model covering all 24 scenarios, while NeRF$^2$ and MLP require per-scene training, totaling 187.2 minutes and 96.7 minutes respectively across 24 scenes. More importantly, when adapting to a dynamic scene, NeRF$^2$ and MLP must train dedicated models from scratch, requiring 7.8 and 4.03 minutes respectively, whereas RadTwin requires no additional training. For inference, NeRF$^2$ requires 8.5 ms per sample due to the computational overhead of volume rendering along each ray, RadTwin achieves 0.61 ms through its efficient Transformer-based architecture, and MLP is the fastest at 0.07 ms owing to its simple structure.

\begin{table}[t]
\centering
\caption{Computational Time Comparison}
\label{tab:time_comparison}
\begin{tabular}{lccc}
\toprule
Method & \makecell{Training Time \\ (per scene)} & \makecell{Adaptation \\ (new scene)} & \makecell{Inference \\ (per sample)} \\
\midrule
RadTwin & -- & 0 s & 0.61 ms \\
NeRF$^2$ & 7.8 min & 7.8 min & 8.5 ms \\
MLP & 4.03 min & 4.03 min & 0.07 ms \\
\bottomrule
\end{tabular}
\end{table}

\subsection{Impact of Training Data Granularity}
\label{subsec:granularity}
\begin{figure}[!t]
	\centering
	\includegraphics[width=3.3in]{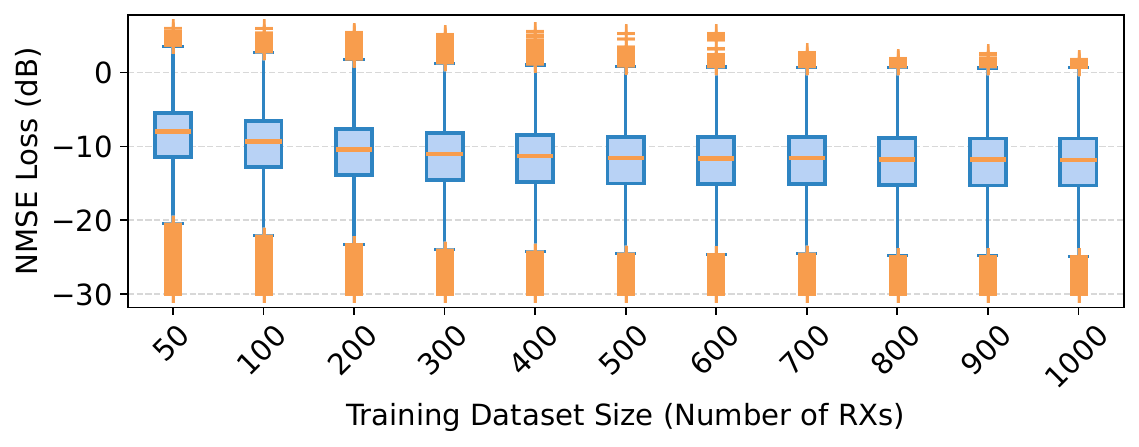}
	\caption{NMSE distribution under varying training data sizes.}
	\label{fig:data_size}
\end{figure}
We investigate how training data volume affects prediction performance by varying the number of RX positions used for training. For each of the 24 training scenes, we sample 1,000 RX positions but use only a subset for training, ranging from 50 to 1,000. All models are evaluated on 6 test scenes using all 1,000 RX positions per scene.

Fig.~\ref{fig:data_size} shows the NMSE distribution across different training data sizes. RadTwin demonstrates consistent improvement as training data increases, with mean NMSE improving from $-8.98$\,dB at 50 RX positions to $-12.64$\,dB at 1,000 RX positions. The improvement is most pronounced below 400 RXs, where mean NMSE drops by 3.17\,dB (from $-8.98$\,dB to $-12.15$\,dB). Beyond 600 RXs, performance gains saturate, with mean NMSE remaining around $-12.5$\,dB. This saturation suggests that approximately 600 RXs per scene provide sufficient spatial coverage for small scenes, and additional measurements yield diminishing returns. These results indicate that RadTwin can achieve high data efficiency and maintain strong generalization performance even with relatively limited training coverage.

% \begin{table*}[htbp]
% \small
% \centering
% \caption{NMSE loss statistics (dB) comparison under different training RX numbers}
% \label{tab:nmse_statistics}
% \begin{tabular}{lcccc|cccc|cccc}
% \toprule
% \multirow{2}{*}{\textbf{RXs}} & \multicolumn{4}{c}{\textbf{RadTwin}} & \multicolumn{4}{c}{\textbf{MLP}} & \multicolumn{4}{c}{\textbf{NeRF$^2$}} \\
% & Mean & Std & Min & Max & Mean & Std & Min & Max & Mean & Std & Min & Max \\
% \midrule
% 100  & -10.20 & 5.31 & -80.00 & 5.97  & -0.21 & 0.06 & -0.68  & -0.12 & -9.77  & 5.08 & -71.45 & 3.91 \\
% 200  & -11.25 & 5.35 & -80.00 & 5.43  & -0.40 & 0.12 & -1.40  & -0.24 & -9.84  & 5.14 & -70.19 & 4.00 \\
% 300  & -11.88 & 5.36 & -80.00 & 5.20  & -0.61 & 0.19 & -2.27  & -0.36 & -9.93  & 5.15 & -66.64 & 3.86 \\
% 400  & -12.15 & 5.37 & -80.00 & 5.53  & -0.83 & 0.27 & -3.35  & -0.48 & -9.47  & 5.18 & -63.36 & 4.17 \\
% 500  & -12.40 & 5.37 & -80.00 & 5.25  & -1.06 & 0.36 & -4.79  & -0.60 & -9.66  & 5.17 & -69.12 & 4.15 \\
% 600  & -12.46 & 5.39 & -80.00 & 5.31  & -1.30 & 0.46 & -6.96  & -0.73 & -10.00 & 5.13 & -67.88 & 3.98 \\
% 800  & -12.58 & 5.43 & -80.00 & 1.91  & -1.85 & 0.74 & -17.72 & -1.00 & -10.02 & 5.12 & -68.81 & 4.14 \\
% 1000 & -12.64 & 5.40 & -69.08 & 1.81  & -2.52 & 1.23 & -37.77 & -1.28 & -10.03 & 5.12 & -80.00 & 4.14 \\
% \bottomrule
% \end{tabular}
% \end{table*}

% TODO: Add experimental results and analysis
% Test configurations: 100, 300, 500, 700, 900 RX positions per scene
% Compare: RadTwin, NeRF2, MLP

\subsection{Scalability to Different Scene Sizes}
\label{subsec:scalability}
We evaluate RadTwin's scalability across the three scene sizes described in Section~\ref{subsec:dataset}. All three sizes use 400 RXs for training. The voxel size is kept constant at $0.5 \times 0.5 \times 0.5$\,m$^3$, resulting in increasing numbers of voxels for larger scenes.

Fig.~\ref{fig:scalability} shows the SNR CDF curves for each scene size. The small scene achieves the highest median SNR of 11.36\,dB due to its simpler propagation environment with fewer multipath components. As scene size increases, the propagation environment becomes more complex with longer path lengths and more potential reflectors. The performance gap also reflects the decreasing spatial density of training samples, as the same 400 RXs provide denser coverage for smaller scenes but become increasingly sparse for larger environments. Nevertheless, RadTwin maintains robust performance with median SNR of 10.79\,dB for the medium scene and 10.12\,dB for the large scene, showing that our voxel-based representation and physics-informed attention mechanism scale effectively to larger indoor environments even with relatively sparse training coverage.

% \begin{figure}[!t]
% 	\centering
% 	\includegraphics[width=3.3in]{fig/scalability.pdf}
% 	\caption{Performance comparison across different scene sizes.}
% 	\label{fig:scalability}
% \end{figure}

\subsection{Generalization to Dynamic Scene Variations}
\label{subsec:dynamic}
A key capability of RadTwin is adapting to dynamic scene configurations without retraining. To evaluate this, we generate 30 scene snapshots by progressively moving furniture from one configuration to another, simulating continuous environmental changes. We vary training set diversity by controlling the sampling step: step = 1 uses all 24 scenes, step = 2 uses 12 scenes, step = 4 uses 6 scenes, step = 8 uses 3 scenes, and step = 24 uses only 1 scene at the extremes. All models are tested on 6 test scenes.

\begin{figure}[!t]
\captionsetup{justification=centering}
  \begin{minipage}[t]{0.24\textwidth}
    \centering
    \includegraphics[width=1.7in]{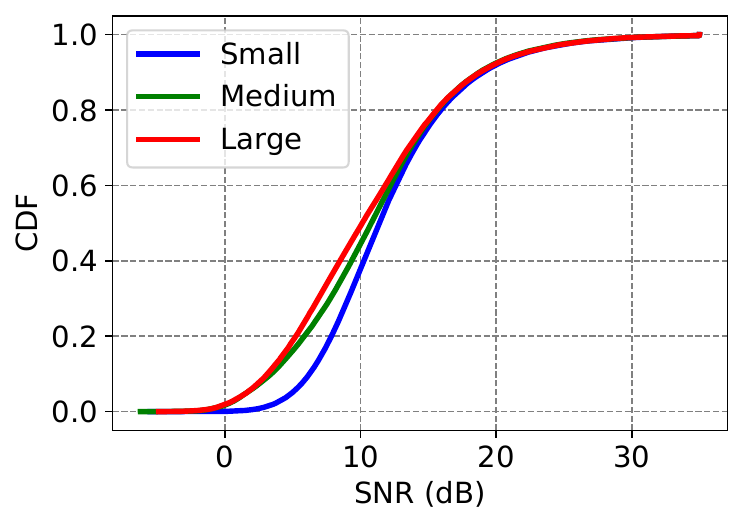}
    \caption{SNR distribution across different scene sizes.}
    \label{fig:scalability}
  \end{minipage}
  \begin{minipage}[t]{0.24\textwidth}
    \centering
    \includegraphics[width=1.7in]{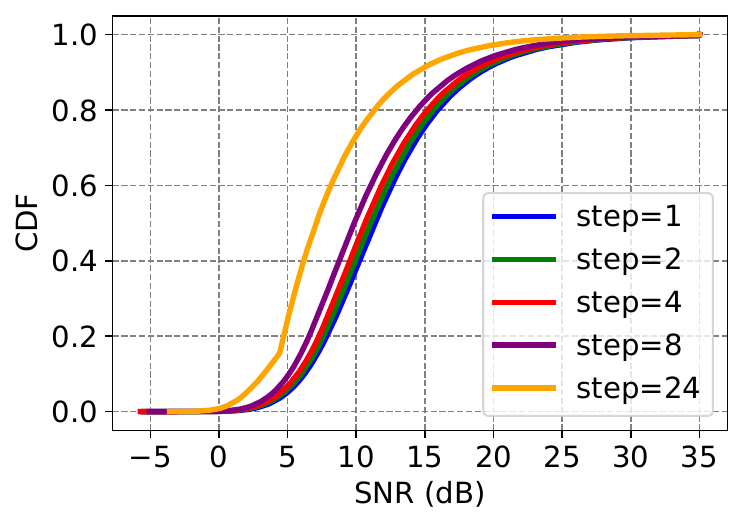}
    \caption{SNR distribution under varying training set diversity.}
    \label{fig:dynamic}
  \end{minipage}
\end{figure}

Fig.~\ref{fig:dynamic} shows the SNR CDF under different training set diversities. RadTwin maintains strong performance even with limited training coverage. Using all 30 scenes achieves a median SNR of 11.36\,dB, while step = 4 with only 8 training scenes still achieves 10.63\,dB. Even step = 8 with 4 training scenes maintains a median SNR of 9.85\,dB. Performance degrades more noticeably at step = 24 with only 2 training scenes, achieving a median SNR of 7.07\,dB. These results demonstrate that RadTwin's explicit geometric conditioning enables effective interpolation to intermediate furniture positions not present in the training set, which is critical for practical deployment in dynamic indoor environments.

\section{Related Work}
\label{sec:related}
\textbf{Simulator-based Approaches.}
Traditional wireless channel modeling relies on either statistical or deterministic methods. Statistical models characterize fading and shadowing using probability distributions~\cite{sarkar2003survey}, while empirical models such as Okumura-Hata~\cite{hata2013empirical} and COST action models~\cite{liu2012cost} derive path loss formulas from extensive measurement campaigns. Standardized geometry-based stochastic models from 3GPP and ITU provide parameterized multipath clustering models for system-level simulations~\cite{zhu20213gpp,series2009guidelines}. However, these approaches fail to capture site-specific propagation characteristics as they approximate path loss as radially symmetric functions of distance.
Deterministic ray tracing methods simulate physical wave propagation by tracing ray paths and computing interactions with environmental obstacles~\cite{yun2015ray}. Modern tools such as Sionna RT~\cite{hoydis2023sionna} provide accurate channel predictions by modeling reflection, diffraction, and scattering and further support differentiable ray tracing for gradient-based optimization of material properties and antenna configurations. However, ray tracing suffers from prohibitive computational complexity and requires precise 3D models with detailed material properties (e.g., permittivity and conductivity), making it impractical for real-time applications in dynamic environments where scene geometry frequently changes.

\textbf{Neural Network-based Approaches.}
Recent deep learning methods have shown promise in wireless channel modeling by learning complex propagation patterns directly from data. RadioUNet~\cite{levie2021radiounet} demonstrates the effectiveness of CNNs for predicting radio maps from 2D urban environment representations, achieving fast inference through the U-Net architecture. However, CNNs struggle to capture long-range spatial dependencies and are limited to 2D representations without modeling 3D furniture-level variations.
Inspired by NeRF~\cite{mildenhall2021nerf}, several works extend the concept to wireless channels. NeRF$^2$~\cite{zhao2023nerf2} pioneered neural RF radiance fields for spatial spectrum synthesis by representing the scene as a continuous volumetric function learned through MLPs. NeWRF~\cite{lu2024newrf} extends this framework for channel prediction from sparse measurements by incorporating wireless propagation physics. WRF-GS~\cite{wen2025neural} and RF-3DGS~\cite{zhang2024rf} adapt 3D Gaussian splatting for faster rendering, achieving millisecond-level inference while maintaining competitive accuracy. However, these methods encode scene geometry implicitly within network weights, necessitating complete model retraining whenever the scenario changes. The learnable wireless digital twin framework~\cite{jiang2025learnable} represents a notable step toward generalizability by combining geometric ray tracing with neural modules to learn EM properties and interaction behaviors of objects. However, it requires accurate 3D mesh models as input, which are costly to obtain and difficult to maintain in dynamic settings.

% mimic

\section{Conclusion}
\label{sec:conclusion}
In this paper, we presented RadTwin, a generalizable DNT framework for wireless channel prediction in dynamic indoor environments. RadTwin achieves explicit geometric conditioning through point cloud-based voxel representation, physics-informed sparse attention via electromagnetic ray tracing, and spatial spectrum prediction through masked cross-attention aggregation. Extensive experiments on 30 indoor scenes demonstrate that RadTwin substantially outperforms state-of-the-art methods with 31.6\% higher SSIM and 91.96\% lower LPIPS, while maintaining robust cross-scale performance and effective adaptation to dynamic configurations without retraining. 
% Our current evaluation relies on clean, uniformly sampled point clouds. In practice, point clouds captured by LiDAR/depth cameras are subject to artifacts such as sparsity, occlusion, and measurement noise.

\section*{Acknowledgement}
This work is partially supported by the US National Science Foundation under Grant No. 2321699 and No. 2333164. 
We appreciate the hardware support from the NVIDIA Academic Grant Award.

\bibliographystyle{IEEEtran}
\bibliography{ref/reference}

\end{document}